\documentclass[twocolumn]{revtex4}
\usepackage{amsmath}

\begin{document}

\title{ Symplectic-dilation mixed wavelet transform and its correspondence
in quantum optics}
\author{Hong-yi Fan$^{1,2},$ Shu-guang Liu$^{1}$ and Li-yun Hu$^{2\ast }$}
\affiliation{$^{1}$Department of Material Science and Engineering,\\
University of Science and Technology of China,\ Hefei, Anhui 230026, China\\
$^{2}$Department of Physics, Shanghai Jiao Tong University, Shanghai 200030,
China\\
$^{\ast }$ Corresponding author. hlyun@sjtu.edu.cn.}

\begin{abstract}
{\small The symplectic wavelet transformation [Opt. Lett. 31 (2006) 3432],
which is related to quantum optical Fresnel transform, is developed to the
symplectic-dilation mixed wavelet transform (SDWT). The SDWT involves both a
real-variable dilation-transform\ and a complex-variable symplectic
transform, and possesses well-behaved properties such as Parseval theorem
and inversion formula. The entangled-coherent state representation (ESCR)
not only underlies the SDWT, but also helps to derive the corresponding
quantum transform operator whose counterpart in classical optics is the
lens-Fresnel mixed transform [Phys. Lett. A 357 (2006) 163].}

{\small OCIS codes: 070.6020, 000.3860.}
\end{abstract}

\maketitle

In recent years wavelet transform (WT) \cite{r1,r2,r3,r4,r5} has been
developed which can overcome the shortcomings of classical Fourier analysis
and it is now recognized as a very useful tool in signal and image
processing. A wavelet has its energy concentrated in time to give a tool for
the analysis of transient, nonstationary, or time-varying phenomena (it is a
wavelet because it is localized and it resembles a wave because it
oscillates). The 1-dimensional continuous WT of a signal function $f\left(
x\right) \in $ $L^{2}\left( R\right) $ by a mother wavelet $\varphi \left(
x\right) $ (restricted by the admissibility condition $\int_{-\infty
}^{\infty }\varphi \left( x\right) \mathtt{d}x=0$) is defined by%
\begin{equation}
W_{\varphi }f\left( \mathrm{a},b\right) =\frac{1}{\sqrt{\left\vert \mathrm{a}%
\right\vert }}\int_{-\infty }^{\infty }f\left( x\right) \varphi ^{\ast
}\left( \frac{x-b}{\mathrm{a}}\right) \mathtt{d}x.  \label{1}
\end{equation}%
($0\neq \mathrm{a}\in R,$ $b\in R$). Dirac's coordinate representation $%
\left\vert x\right\rangle $ in quantum mechanics underlies the WT in (\ref{1}%
), since (\ref{1}) can be recast as $W_{\varphi }f\left( \mathrm{a},b\right)
=\left\langle \varphi \right\vert U^{\left( \mathrm{a},b\right) }\left\vert
f\right\rangle \ $\cite{r6}, where $\left\langle \varphi \right\vert $ is
the bra vector corresponding to the given mother wavelet, $\left\vert
f\right\rangle $ is the state to be transformed, $f\left( x\right)
=\left\langle x\right\vert \left. f\right\rangle ,$ and $U^{\left( \mathrm{a}%
,b\right) }=\frac{1}{\sqrt{\left\vert \mathrm{a}\right\vert }}\int_{-\infty
}^{\infty }\left\vert \frac{x-b}{\mathrm{a}}\right\rangle \left\langle
x\right\vert \mathtt{d}x,$ is a dilated-translated operator, $U^{\left(
\mathrm{a},0\right) }=\exp [\frac{1}{2}(a_{1}^{\dagger 2}-a_{1}^{2})\ln
\mathrm{a}]$ (for $\mathrm{a}>0$) is the single-mode squeezing operator \cite%
{r7a,r7b}. Motivated by the consideration that the symplectic transforms $%
\left(
\begin{array}{cc}
s & -r \\
-r^{\ast } & -s^{\ast }%
\end{array}%
\right) ,$ where $s,r$ are both complex parameters, $\left\vert s\right\vert
^{2}-\left\vert r\right\vert ^{2}=1$, are more general than the squeezing
transform, in Ref. \cite{r7} Fan and Lu proposed the symplectic-transformed
translated wavelet family $\phi _{r,s}^{\ast }\left( z-\kappa \right) $
generated from the mother wavelet $\phi \left( z\right) _{z=z_{1}+\mathtt{i}%
z_{2}}$ and the symplectic WT (SWT) for a complex function $f\left( z\right)
,$%
\begin{equation}
W_{\phi }f\left( r,s;\kappa \right) =\int \frac{\mathtt{d}^{2}z}{\pi }%
f\left( z\right) \phi _{r,s}^{\ast }\left( z-\kappa \right) ,\text{ \ \ }%
\mathtt{d}^{2}z=\mathtt{d}z_{1}\mathtt{d}z_{2},  \label{4}
\end{equation}%
where $\kappa $ is a complex translation parameter, $\phi _{r,s}\left(
z-\kappa \right) \equiv \sqrt{s^{\ast }}\phi \left[ s\left( z-\kappa \right)
-r\left( z^{\ast }-\kappa ^{\ast }\right) \right] .$ The quantum mechanical
representation that underlies the SWT is the coherent state \cite{r8,r9,r10}%
, $\left\vert z\right\rangle =\exp \left( za_{1}^{\dagger }-z^{\ast
}a_{1}\right) \left\vert 0\right\rangle \equiv \left\vert \binom{z}{z^{\ast }%
}\right\rangle ,$ possessing the completeness relation $\int \frac{\mathtt{d}%
^{2}z}{\pi }\left\vert z\right\rangle \left\langle z\right\vert =1$. The
quantum mechanical version of (\ref{4}) for $f\left( z\right) \equiv
\left\langle z\right\vert \left. f\right\rangle \ $and $\phi _{r,s}\left(
z-\kappa \right) \equiv \left\langle s\left( z-\kappa \right) -r\left(
z^{\ast }-\kappa ^{\ast }\right) \right. \left\vert \phi \right\rangle ,$ is%
\begin{eqnarray}
W_{\phi }f\left( r,s;k\right) &=&\sqrt{s}\int \frac{\mathtt{d}^{2}z}{\pi }%
\left\langle \phi \right. \left\vert s\left( z-\kappa \right) -r\left(
z^{\ast }-\kappa ^{\ast }\right) \right\rangle \left\langle z\right.
\left\vert f\right\rangle  \notag \\
&=&\left\langle \phi \right\vert F^{(r,s,\kappa )}\left\vert f\right\rangle ,
\label{7}
\end{eqnarray}%
where $F^{\left( r,s;\kappa \right) }$ is defined as $F^{\left( r,s;\kappa
\right) }=\sqrt{s}\int \frac{\mathtt{d}^{2}z}{\pi }\left\vert sz-rz^{\ast
}\right\rangle \left\langle z+\kappa \right\vert ,$ $\left\vert sz-rz^{\ast
}\right\rangle \equiv \left\vert \left(
\begin{array}{cc}
s & -r \\
-r^{\ast } & -s^{\ast }%
\end{array}%
\right) \left(
\begin{array}{c}
z \\
z^{\ast }%
\end{array}%
\right) \right\rangle .$ So the SWT is performed over the mother wavelet
gained through a translating transform followed by a symplectic transform.
The SWT is worth mentioning because $F^{\left( r,s;\kappa =0\right) }=e^{-%
\frac{r}{2s^{\ast }}a_{1}^{\dagger 2}}e^{(a_{1}^{\dagger }a_{1}+\frac{1}{2}%
)\ln \frac{1}{s^{\ast }}}e^{\frac{r^{\ast }}{2s^{\ast }}a_{1}^{2}}$ is just
the single-mode Fresnel operator \cite{r11} corresponding to the classical
Fresnel optical transform \cite{r11a}.

Recalling that in Ref. \cite{r12} we have constructed a new
entangled-coherent state representation $\left\vert \alpha ,x\right\rangle $,%
\begin{eqnarray}
\left\vert \alpha ,x\right\rangle &=&\exp \left[ -\frac{1}{2}x^{2}-\frac{1}{4%
}\left\vert \alpha \right\vert ^{2}+(x+\frac{\alpha }{2})a_{1}^{\dagger
}\right.  \notag \\
&&\left. +(x-\frac{\alpha }{2})a_{2}^{\dagger }-\frac{1}{4}(a_{1}^{\dagger
}+a_{2}^{\dagger })^{2}\right] \left\vert 00\right\rangle ,  \label{10}
\end{eqnarray}%
which is the common eigenvector of the operator $\left( X_{1}+X_{2}\right)
/2 $ and $a_{1}-a_{2},$ i.e., $\left( a_{1}-a_{2}\right) \left\vert \alpha
,x\right\rangle =\alpha \left\vert \alpha ,x\right\rangle $ and $\frac{1}{2}%
(X_{1}+X_{2})\left\vert \alpha ,x\right\rangle =\frac{1}{\sqrt{2}}%
x\left\vert \alpha ,x\right\rangle ,$ where $X_{i}=\frac{1}{\sqrt{2}}%
(a_{i}+a_{i}^{\dagger })$ is coordinate operator, ($i=1,2)$. $\left\vert
\alpha ,x\right\rangle $ constitutes a complete representation,%
\begin{equation}
\int_{-\infty }^{\infty }\frac{\mathtt{d}x}{\sqrt{\pi }}\int \frac{\mathtt{d}%
^{2}\alpha }{2\pi }\left\vert \alpha ,x\right\rangle \left\langle \alpha
,x\right\vert =1,  \label{12}
\end{equation}%
and exhibits partly non-orthogonal property (for $\alpha )$ and
Delta-function orthonormal property (for $x),$
\begin{eqnarray}
&&\left\langle \alpha ^{\prime },x^{\prime }\right. \left\vert \alpha
,x\right\rangle  \notag \\
&=&\sqrt{\pi }\exp \left[ -\frac{1}{4}(\left\vert \alpha \right\vert
^{2}+\left\vert \alpha ^{\prime }\right\vert ^{2})+\frac{1}{2}\alpha \alpha
^{\prime \ast }\right] \delta \left( x^{\prime }-x\right) .  \label{13}
\end{eqnarray}%
so $\left\vert \alpha ,x\right\rangle $ possess behaviour of both the
coherent state and the entangled state. An interesting question thus
naturally arises: Can we introduce a new kind of continuous WT for which the
$\left\vert \alpha ,x\right\rangle $ representation underlies? The answer is
affirmative. Our motivation of this issue comes from the mixed lens-Fresnel
transform in classical optics \cite{r13} (see (\ref{42}) and (\ref{44})
below).

By synthesizing (\ref{1}) and (\ref{4}) and in reference to (\ref{12}) we
propose the mixed WT for $g\left( \alpha ,x\right) $ ($\alpha =\alpha _{1}+%
\mathtt{i}\alpha _{2}\ $is a complex and is independent of $x$):
\begin{equation}
W_{\psi }g\left( s,r,\kappa ;\mathrm{a},b\right) \equiv \int_{-\infty
}^{\infty }\frac{\mathtt{d}x}{\sqrt{\pi }}\int \frac{\mathtt{d}^{2}\alpha }{%
2\pi }g\left( \alpha ,x\right) \psi _{s,r,\kappa ;\mathrm{a},b}^{\ast
}\left( \alpha ,x\right) .  \label{14}
\end{equation}%
where $\mathtt{d}^{2}\alpha =\mathtt{d}\alpha _{1}\mathtt{d}\alpha _{2},$
the family of mother wavelet $\psi $ involves both the the symplectic
transform of $\alpha $ and the dilation-transform\ of $x$,%
\begin{equation}
\psi _{s,r,\kappa ;\mathrm{a},b}\left( \alpha ,x\right) =\sqrt{\frac{s^{\ast
}}{\left\vert \mathrm{a}\right\vert }}\psi \left[ s\left( \alpha -\kappa
\right) -r\left( \alpha ^{\ast }-\kappa ^{\ast }\right) ,\frac{x-b}{\mathrm{a%
}}\right] .  \label{15}
\end{equation}%
Letting $g\left( \alpha ,x\right) \equiv $ $\left\langle \alpha
,x\right\vert \left. g\right\rangle ,$ the wave function of state vector $%
\left\vert g\right\rangle $ in the $\left\langle \alpha ,x\right\vert $
representation, then Eq. (\ref{14}) can be expressed as quantum mechanical
version%
\begin{eqnarray}
&&W_{\psi }g\left( s,r,\kappa ;\mathrm{a},b\right)   \notag \\
&=&\sqrt{\frac{s}{\left\vert \mathrm{a}\right\vert }}\int_{-\infty }^{\infty
}\frac{\mathtt{d}x}{\sqrt{\pi }}\int \frac{\mathtt{d}^{2}\alpha }{2\pi }%
g\left( \alpha ,x\right)   \notag \\
&&\times \psi ^{\ast }\left[ s\left( \alpha -\kappa \right) -r\left( \alpha
^{\ast }-\kappa ^{\ast }\right) ,\frac{x-b}{\mathrm{a}}\right]   \notag \\
&=&\left\langle \psi \right\vert U\left( s,r,\kappa ;\mathrm{a},b\right)
\left\vert g\right\rangle ,  \label{16}
\end{eqnarray}%
where $U\left( s,r,\kappa ;\mathrm{a},b\right) $ is defined as%
\begin{eqnarray}
U\left( s,r,\kappa ;\mathrm{a},b\right)  &=&\sqrt{\frac{s}{\left\vert
\mathrm{a}\right\vert }}\int_{-\infty }^{\infty }\frac{\mathtt{d}x}{\sqrt{%
\pi }}\int \frac{\mathtt{d}^{2}\alpha }{2\pi }  \notag \\
&&\times \left\vert s\alpha -r\alpha ^{\ast },\frac{x-b}{\mathrm{a}}%
\right\rangle \left\langle \alpha +\kappa ,x\right\vert .  \label{17}
\end{eqnarray}%
$U\left( s,r,\kappa =0;\mathrm{a},b=0\right) $ is just the generalized
squeezing operator $U\left( s,r,\mathrm{a}\right) $ $($see (\ref{35})
below), which causes a lens-Fresnel mixed transform.

For Eq. (\ref{14}) being qualified as a new WT we must prove that it
possesses fundamental properties of the usual WTs, such as the admissibility
condition, the Parseval theorem and the inversion formula. It is
straightforward to evaluate the transform (\ref{14}) and its reciprocal
transform when $g\left( \alpha ,x\right) $\ is the exponential $g_{1}\left(
\alpha ,x\right) =\exp \left( \alpha ^{\ast }\beta -\alpha \beta ^{\ast }-%
\mathtt{i}px\right) ,$%
\begin{eqnarray}
W_{\psi }g_{1} &=&\int_{-\infty }^{\infty }\frac{\mathtt{d}x}{\sqrt{\pi }}%
\int \frac{\mathtt{d}^{2}\alpha }{2\pi }\psi _{s,r,\kappa ;\mathrm{a}%
,b}^{\ast }\left( \alpha ,x\right) e^{\alpha ^{\ast }\beta -\alpha \beta
^{\ast }-\mathtt{i}px}  \notag \\
&=&\sqrt{\frac{s}{\left\vert \mathrm{a}\right\vert }}e^{\kappa ^{\ast }\beta
-\kappa \beta ^{\ast }-\mathtt{i}pb}\int_{-\infty }^{\infty }\frac{\mathtt{d}%
x}{\sqrt{\pi }}\int \frac{\mathtt{d}^{2}\alpha }{2\pi }  \notag \\
&&\times \psi ^{\ast }\left[ s\alpha -r\alpha ^{\ast },\frac{x}{\mathrm{a}}%
\right] e^{\alpha ^{\ast }\beta -\alpha \beta ^{\ast }-\mathtt{i}px}.
\label{18}
\end{eqnarray}%
Making the integration variables transform $s\alpha -r\alpha ^{\ast
}\rightarrow w,\frac{x}{\mathrm{a}}\rightarrow x^{\prime },$ leading to $%
\mathtt{d}^{2}\alpha \rightarrow \mathtt{d}^{2}w$ and $\int_{-\infty
}^{\infty }\mathtt{d}x\rightarrow \left\vert \mathrm{a}\right\vert
\int_{-\infty }^{\infty }\mathtt{d}x^{\prime }$, Eq. (\ref{18}) becomes%
\begin{equation}
W_{\psi }g_{1}=\sqrt{s\left\vert \mathrm{a}\right\vert }\Phi ^{\ast }\left(
s^{\ast }\beta ^{\ast }-r^{\ast }\beta ,\text{ }\mathrm{a}p\right) e^{\kappa
^{\ast }\beta -\kappa \beta ^{\ast }-\mathtt{i}pb}.  \label{19}
\end{equation}%
where $\Phi ^{\ast }$ is just the Fourier transform of $\psi ^{\ast },$%
\begin{eqnarray}
\Phi ^{\ast }\left( s^{\ast }\beta ^{\ast }-r^{\ast }\beta ,\text{ }\mathrm{a%
}p\right) &=&\int_{-\infty }^{\infty }\frac{\mathtt{d}x^{\prime }}{\sqrt{\pi
}}\int \frac{\mathtt{d}^{2}w}{2\pi }\psi ^{\ast }\left[ w,x^{\prime }\right]
\notag \\
&&\times e^{w^{\ast }\left( s\beta -r\beta ^{\ast }\right) -w\left( s^{\ast
}\beta ^{\ast }-r^{\ast }\beta \right) -\mathtt{i}\mathrm{a}px^{\prime }}.
\label{20}
\end{eqnarray}%
Then we perform the adjoint WT of (\ref{14}), using (\ref{15})\ and (\ref{19}%
) we see%
\begin{eqnarray}
&&W_{\psi }^{\ast }\left( W_{\psi }g_{1}\right) \left( \alpha ,x\right)
\notag \\
&=&\sqrt{\frac{s^{\ast }}{\left\vert \mathrm{a}\right\vert }}\int_{-\infty
}^{\infty }\frac{\mathtt{d}b}{\sqrt{\pi }}\int \frac{\mathtt{d}^{2}\kappa }{%
2\pi }W_{\psi }g_{1}  \notag \\
&&\times \psi \left[ s\left( \alpha -\kappa \right) -r\left( \alpha ^{\ast
}-\kappa ^{\ast }\right) ,\frac{x-b}{\mathrm{a}}\right]  \notag \\
&=&\left\vert s\right\vert \Phi ^{\ast }\left( s^{\ast }\beta ^{\ast
}-r^{\ast }\beta ,\text{ }\mathrm{a}p\right) \int_{-\infty }^{\infty }\frac{%
\mathtt{d}b}{\sqrt{\pi }}\int \frac{\mathtt{d}^{2}\kappa }{2\pi }  \notag \\
&&\times e^{\kappa ^{\ast }\beta -\kappa \beta ^{\ast }-\mathtt{i}pb}\psi %
\left[ s\left( \alpha -\kappa \right) -r\left( \alpha ^{\ast }-\kappa ^{\ast
}\right) ,\frac{x-b}{\mathrm{a}}\right]  \notag \\
&=&\left\vert s\right\vert \left\vert \mathrm{a}\right\vert g_{1}\left(
\alpha ,x\right) \Phi ^{\ast }\left( s^{\ast }\beta ^{\ast }-r^{\ast }\beta ,%
\text{ }\mathrm{a}p\right) \int_{-\infty }^{\infty }\frac{\mathtt{d}%
b^{\prime }}{\sqrt{\pi }}  \notag \\
&&\times \int \frac{\mathtt{d}^{2}\kappa ^{\prime }}{2\pi }e^{\kappa
^{\prime }\beta ^{\ast }-\kappa ^{\prime \ast }\beta +\mathtt{i}\mathrm{a}%
pb^{\prime }}\psi \left[ s\kappa ^{\prime }-r\kappa ^{\prime \ast
},b^{\prime }\right]  \notag \\
&=&\left\vert s\right\vert \left\vert \mathrm{a}\right\vert g_{1}\left(
\alpha ,x\right) \left\vert \Phi \left( s^{\ast }\beta ^{\ast }-r^{\ast
}\beta ,\text{ }\mathrm{a}p\right) \right\vert ^{2}.  \label{21}
\end{eqnarray}%
From Eq. (\ref{21}) we obtain%
\begin{eqnarray}
&&\int_{-\infty }^{\infty }\frac{\mathtt{d}\mathrm{a}}{\mathrm{a}^{2}}\int
\frac{\mathtt{d}^{2}s}{\left\vert s\right\vert ^{2}}W_{\psi }^{\ast }\left(
W_{\psi }g_{1}\right) \left( \alpha ,x\right)  \notag \\
&=&g_{1}\left( \alpha ,x\right) \int_{-\infty }^{\infty }\frac{\mathtt{d}%
\mathrm{a}}{\left\vert \mathrm{a}\right\vert }\int \frac{\mathtt{d}^{2}s}{%
\left\vert s\right\vert }\left\vert \Phi \left( s^{\ast }\beta ^{\ast
}-r^{\ast }\beta ,\text{ }\mathrm{a}p\right) \right\vert ^{2},  \label{22}
\end{eqnarray}%
which leads to
\begin{equation}
g_{1}\left( \alpha ,x\right) =\frac{\int_{-\infty }^{\infty }\frac{\mathtt{d}%
a}{a^{2}}\int \frac{\mathtt{d}^{2}s}{\left\vert s\right\vert ^{2}}W_{\psi
}^{\ast }\left( W_{\psi }g_{1}\right) \left( \alpha ,x\right) }{%
\int_{-\infty }^{\infty }\frac{\mathtt{d}a}{\left\vert a\right\vert }\int
\frac{\mathtt{d}^{2}s}{\left\vert s\right\vert }\left\vert \Phi \left(
s^{\ast }\beta ^{\ast }-r^{\ast }\beta ,\text{ }ap\right) \right\vert ^{2}}.
\label{23}
\end{equation}%
Eq. (\ref{23}) implies that we should impose the normalization
\begin{equation}
\int_{-\infty }^{\infty }\frac{\mathtt{d}\mathrm{a}}{\left\vert \mathrm{a}%
\right\vert }\int \frac{\mathtt{d}^{2}s}{\left\vert s\right\vert }\left\vert
\Phi \left( s^{\ast }\beta ^{\ast }-r^{\ast }\beta ,\text{ }\mathrm{a}%
p\right) \right\vert ^{2}=1,  \label{24}
\end{equation}%
such that the reproducing process\textit{\ }exists
\begin{equation}
g_{1}\left( \alpha ,x\right) =\int_{-\infty }^{\infty }\frac{\mathtt{d}%
\mathrm{a}}{\mathrm{a}^{2}}\int \frac{\mathtt{d}^{2}s}{\left\vert
s\right\vert ^{2}}W_{\psi }^{\ast }\left( W_{\psi }g_{1}\right) \left(
\alpha ,x\right) .  \label{25}
\end{equation}%
(\ref{24}) may be named the generalized \textit{admissibility condition}.
Now we can have the corresponding \emph{Parseval theorem}: For any $g$ and $%
g^{\prime }$ we have%
\begin{eqnarray}
&&\int_{-\infty }^{\infty }\frac{\mathtt{d}\mathrm{a\mathtt{d}}b}{\mathrm{a}%
^{2}}\int \frac{\mathtt{d}^{2}\kappa \mathtt{d}^{2}s}{\left\vert
s\right\vert ^{2}}W_{\psi }g\left( s,r,\kappa ;\mathrm{a},b\right) W_{\psi
}^{\ast }g^{\prime }\left( s,r,\kappa ;\mathrm{a},b\right)  \notag \\
&=&\int_{-\infty }^{\infty }\mathtt{d}x\int \mathtt{d}^{2}\alpha g\left(
\alpha ,x\right) g^{\prime \ast }\left( \alpha ,x\right) .  \label{26}
\end{eqnarray}%
\emph{Proof:}\textbf{\ }Assuming $F\left( \beta ,p\right) $ and $F^{\prime
}\left( \beta ,p\right) $ be the Fourier transform of $g\left( \alpha
,x\right) $ and $g^{\prime }\left( \alpha ,x\right) $, respectively,%
\begin{equation}
F\left( \beta ,p\right) =\int_{-\infty }^{\infty }\frac{\mathtt{d}x}{\sqrt{%
2\pi }}\int \frac{\mathtt{d}^{2}\alpha }{\pi }g\left( \alpha ,x\right)
e^{\alpha \beta ^{\ast }-\alpha ^{\ast }\beta +\mathtt{i}px},  \label{27}
\end{equation}%
whose inversion formula is given by%
\begin{equation}
g\left( \alpha ,x\right) =\int_{-\infty }^{\infty }\frac{\mathtt{d}p}{\sqrt{%
2\pi }}\int \frac{\mathtt{d}^{2}\beta }{\pi }F\left( \beta ,p\right)
e^{\alpha ^{\ast }\beta -\alpha \beta ^{\ast }-\mathtt{i}px}.  \label{28}
\end{equation}%
In order to prove (\ref{26}), we first calculate $W_{\psi }g\left(
s,r,\kappa ;\mathrm{a},b\right) $. In similar to deriving Eq.(\ref{19}),
using (\ref{14}), (\ref{15}) and (\ref{28}) we have%
\begin{eqnarray}
&&W_{\psi }g\left( s,r,\kappa ;\mathrm{a},b\right)  \notag \\
&=&\sqrt{\frac{s}{\left\vert \mathrm{a}\right\vert }}\int_{-\infty }^{\infty
}\frac{\mathtt{d}x}{\sqrt{\pi }}\int \frac{\mathtt{d}^{2}\alpha }{2\pi }%
g\left( \alpha ,x\right)  \notag \\
&&\times \psi ^{\ast }\left[ s\left( \alpha -\kappa \right) -r\left( \alpha
^{\ast }-\kappa ^{\ast }\right) ,\frac{x-b}{\mathrm{a}}\right]  \notag \\
&=&\sqrt{\frac{s}{\left\vert \mathrm{a}\right\vert }}\int_{-\infty }^{\infty
}\frac{\mathtt{d}p}{\sqrt{2\pi }}\int \frac{\mathtt{d}^{2}\beta }{\pi }%
F\left( \beta ,p\right) \int_{-\infty }^{\infty }\frac{\mathtt{d}x}{\sqrt{%
\pi }}\int \frac{\mathtt{d}^{2}\alpha }{2\pi }  \notag \\
&&\times e^{\alpha ^{\ast }\beta -\alpha \beta ^{\ast }-\mathtt{i}px}\psi
^{\ast }\left[ s\left( \alpha -\kappa \right) -r\left( \alpha ^{\ast
}-\kappa ^{\ast }\right) ,\frac{x-b}{\mathrm{a}}\right]  \notag \\
&=&\sqrt{s\left\vert \mathrm{a}\right\vert }\int_{-\infty }^{\infty }\frac{%
\mathtt{d}p}{\sqrt{2\pi }}\int \frac{\mathtt{d}^{2}\beta }{\pi }F\left(
\beta ,p\right)  \notag \\
&&\times \Phi ^{\ast }\left( s^{\ast }\beta ^{\ast }-r^{\ast }\beta ,\text{ }%
\mathrm{a}p\right) e^{\kappa ^{\ast }\beta -\kappa \beta ^{\ast }-\mathtt{i}%
pb}.  \label{29}
\end{eqnarray}%
It then follows that
\begin{eqnarray}
&&\int_{-\infty }^{\infty }\mathtt{d}b\int \mathtt{d}^{2}\kappa W_{\psi
}g\left( s,r,\kappa ;\mathrm{a},b\right) W_{\psi }^{\ast }g^{\prime }\left(
s,r,\kappa ;\mathrm{a},b\right)  \notag \\
&=&\left\vert s\right\vert \left\vert \mathrm{a}\right\vert \int_{-\infty
}^{\infty }\mathtt{d}p\mathtt{d}p^{\prime }\int \mathtt{d}^{2}\beta \mathtt{d%
}^{2}\beta ^{\prime }F\left( \beta ,p\right) F^{\prime \ast }\left( \beta
^{\prime },p^{\prime }\right)  \notag \\
&&\times \Phi ^{\ast }\left( s^{\ast }\beta ^{\ast }-r^{\ast }\beta ,\text{ }%
\mathrm{a}p\right) \Phi \left( s^{\ast }\beta ^{\prime \ast }-r^{\ast }\beta
^{\prime },\mathrm{a}p^{\prime }\right)  \notag \\
&&\times \int_{-\infty }^{\infty }\frac{\mathtt{d}b}{2\pi }\int \frac{%
\mathtt{d}^{2}\kappa }{\pi ^{2}}e^{\kappa ^{\ast }\left( \beta -\beta
^{\prime }\right) -\kappa \left( \beta ^{\ast }-\beta ^{\prime \ast }\right)
+\mathtt{i}\left( p^{\prime }-p\right) b}  \notag \\
&=&\left\vert \mathrm{a}s\right\vert \int_{-\infty }^{\infty }\mathtt{d}%
p\int \mathtt{d}^{2}\beta F\left( \beta ,p\right) F^{\prime \ast }\left(
\beta ,p\right) \left\vert \Phi \left( s^{\ast }\beta ^{\ast }-r^{\ast
}\beta ,\mathrm{a}p\right) \right\vert ^{2}.  \label{30}
\end{eqnarray}%
Substituting (\ref{30}) into the left-hand side (LHS) of (\ref{26}) and
using (\ref{24}) we see%
\begin{eqnarray}
\text{LHS of (\ref{26})} &=&\int_{-\infty }^{\infty }\mathtt{d}p\int \mathtt{%
d}^{2}\beta F\left( \beta ,p\right) F^{\prime \ast }\left( \beta ,p\right)
\notag \\
&&\times \int_{-\infty }^{\infty }\frac{\mathtt{d}\mathrm{a}}{\left\vert
\mathrm{a}\right\vert }\int \frac{\mathtt{d}^{2}s}{\left\vert s\right\vert }%
\left\vert \Phi ^{\ast }\left( s^{\ast }\beta ^{\ast }-r^{\ast }\beta ,%
\mathrm{a}p\right) \right\vert ^{2}  \notag \\
&=&\int_{-\infty }^{\infty }\mathtt{d}p\int \mathtt{d}^{2}\beta F\left(
\beta ,p\right) F^{\prime \ast }\left( \beta ,p\right)  \notag \\
&=&\int_{-\infty }^{\infty }\mathtt{d}x\int \mathtt{d}^{2}\alpha g\left(
\alpha ,x\right) g^{\prime \ast }\left( \alpha ,x\right) .  \label{31}
\end{eqnarray}%
Thus we complete the proof of Eq.(\ref{26}).

\emph{Inversion Formula}:\emph{\ }From Eq. (\ref{26}) we have
\begin{equation}
g\left( \alpha ,x\right) =\int_{-\infty }^{\infty }\frac{\mathtt{d\mathrm{a}d%
}b}{\sqrt{\pi }\mathrm{a}^{2}}\int \frac{\mathtt{d}^{2}\kappa \mathtt{d}^{2}s%
}{2\pi \left\vert s\right\vert ^{2}}W_{\psi }g\left( s,r,\kappa ;\mathrm{a}%
,b\right) \psi _{s,r,\kappa ;\mathrm{a},b}\left( \alpha ,x\right) ,
\label{32}
\end{equation}%
that is the inversion formula for the original signal $g\left( \alpha
,x\right) $ expressed by a superposition of wavelet functions $\psi
_{s,r,\kappa ;\mathrm{a},b}\left( \alpha ,x\right) ,$ with the value of
continuous WT $W_{\psi }g\left( s,r,\kappa ;\mathrm{a},b\right) $ serving as
coefficients. In fact, in Eq. (\ref{14}) when we take $g\left( \alpha
,x\right) =\delta \left( \alpha -\alpha ^{\prime }\right) \delta \left(
\alpha ^{\ast }-\alpha ^{\prime \ast }\right) \delta \left( x-x^{\prime
}\right) ,$ then%
\begin{eqnarray}
W_{\psi }g\left( s,r,\kappa ;\mathrm{a},b\right) &=&\int_{-\infty }^{\infty }%
\frac{\mathtt{d}x}{\sqrt{\pi }}\int \frac{\mathtt{d}^{2}\alpha }{2\pi }\psi
_{s,r,\kappa ;a,b}^{\ast }\left( \alpha ,x\right)  \notag \\
&&\times \delta \left( \alpha -\alpha ^{\prime }\right) \delta \left( \alpha
^{\ast }-\alpha ^{\prime \ast }\right) \delta \left( x-x^{\prime }\right)
\notag \\
&=&\frac{1}{2\pi \sqrt{\pi }}\psi _{s,r,\kappa ;\mathrm{a},b}^{\ast }\left(
\alpha ^{\prime },x^{\prime }\right) .  \label{34}
\end{eqnarray}%
Substituting (\ref{34}) into (\ref{31}) yields (\ref{32}).

We can visualize the new WT $W_{\psi }g\left( s,r,\kappa ;\mathrm{a}%
,b\right) $ in the context of quantum optics. By noticing that the
generalized squeezing operator $U\left( s,r,\kappa =0;a,b=0\right) $ in (\ref%
{17}) is an image of the combined mapping of\ the classical real dilation
transformation $x\rightarrow $ $x/\mathrm{a}$ ($\mathrm{a}>0$) and the
classical complex symplectic transform $\left( \alpha ,\alpha ^{\ast
}\right) \rightarrow \left( s\alpha -r\alpha ^{\ast },s^{\ast }\alpha ^{\ast
}-r^{\ast }\alpha \right) $ in $\left\vert \alpha ,x\right\rangle $
representation, we can use the technique of integration within normal
product of operators \cite{r14,r15} to perform the integration in (\ref{17}%
), the result is

\begin{eqnarray}
&&U\left( s,r,\kappa =0;\mathrm{a},b=0\right)  \notag \\
&=&\frac{\text{sech}{}^{1/2}\lambda }{\sqrt{s^{\ast }}}\exp \left[ -\frac{%
\tanh \lambda }{4}(a_{1}^{\dagger }+a_{2}^{\dagger })^{2}\right.  \notag \\
&&\left. -\frac{r}{4s^{\ast }}(a_{1}^{\dagger }-a_{2}^{\dagger })^{2}\right]
V\left( s,\mathrm{a}\right)  \notag \\
&&\times \exp \left[ \frac{\tanh \lambda }{4}(a_{1}+a_{2})^{2}+\frac{r^{\ast
}}{4s^{\ast }}(a_{1}-a_{2})^{2}\right] ,  \label{35}
\end{eqnarray}%
where $e^{\lambda }=\mathrm{a},$ sech${}\lambda =\frac{2\mathrm{a}}{1+%
\mathrm{a}^{2}},$ $\tanh \lambda =\frac{\mathrm{a}^{2}-1}{1+\mathrm{a}^{2}},$
and%
\begin{eqnarray}
V\left( s,\mathrm{a}\right) &=&\colon \exp \left\{ \left( a_{1}^{\dagger },%
\text{ }a_{2}^{\dagger }\right) \left[ \Lambda -I\right] \binom{a_{1}}{a_{2}}%
\right\} \colon ,\text{ }  \notag \\
\Lambda &\equiv &\frac{1}{2}\left(
\begin{array}{cc}
\text{sech}{}\lambda +\frac{1}{s^{\ast }} & \text{sech}{}\lambda -\frac{1}{%
s^{\ast }} \\
\text{sech}{}\lambda -\frac{1}{s^{\ast }} & \text{sech}{}\lambda +\frac{1}{%
s^{\ast }}%
\end{array}%
\right) .  \label{37}
\end{eqnarray}%
The transformation matrix element of $U\left( s,r,\kappa =0;\mathrm{a}%
,b=0\right) $ in the entangled state representation $\left\vert \eta
\right\rangle $ is \cite{r16}%
\begin{eqnarray}
&&\left\langle \eta \right\vert U\left( s,r,\kappa =0;\mathrm{a},b=0\right)
\left\vert \eta ^{\prime }\right\rangle  \notag \\
&=&\sqrt{\frac{s}{\mathrm{a}}}\int_{-\infty }^{\infty }\frac{\mathtt{d}x}{%
\sqrt{\pi }}\int \frac{\mathtt{d}^{2}\alpha }{2\pi }\left\langle \eta
\right. \left\vert s\alpha -r\alpha ^{\ast },\frac{x}{\mathrm{a}}%
\right\rangle \left\langle \alpha ,x\right\vert \left. \eta ^{\prime
}\right\rangle ,  \label{38}
\end{eqnarray}%
where $\left\vert \eta \right\rangle $ is the bipartite entangled state of
continuous variable. In two-mode Fock space $\left\vert \eta =\eta _{1}+%
\mathtt{i}\eta _{2}\right\rangle $ reads%
\begin{equation}
\left\vert \eta \right\rangle =\exp [-\frac{1}{2}\left\vert \eta \right\vert
^{2}+\eta a_{1}^{\dagger }-\eta ^{\ast }a_{2}^{\dagger }+a_{1}^{\dagger
}a_{2}^{\dagger }]\left\vert 00\right\rangle ,  \label{39}
\end{equation}%
$\left\vert \eta \right\rangle $ is introduced on the concept of quantum
entanglement first proposed in 1935 by Einstein-Podolsky-Rosen (EPR) \cite%
{r17} in the argument that quantum mechanics theory was not complete. Then
using (\ref{10}) and (\ref{39}), we obtain

\begin{equation}
\left\langle \eta \right. \left\vert \alpha ,x\right\rangle =\frac{1}{\sqrt{2%
}}\exp \left[ -\frac{\alpha ^{2}+\left\vert \alpha \right\vert ^{2}}{4}-%
\frac{1}{2}\eta _{1}^{2}+\eta _{1}\alpha -\mathtt{i}\eta _{2}x\right] .
\label{40}
\end{equation}%
It directly follows from (\ref{40}) that$\allowbreak $%
\begin{eqnarray}
\left\langle \eta \right. \left\vert s\alpha -r\alpha ^{\ast },\frac{x}{%
\mathrm{a}}\right\rangle &=&\frac{1}{\sqrt{2}}\exp \left[ -\frac{\left(
s\alpha -r\alpha ^{\ast }\right) ^{2}+\left\vert s\alpha -r\alpha ^{\ast
}\right\vert ^{2}}{4}\right.  \notag \\
&&\left. -\frac{1}{2}\eta _{1}^{2}+\eta _{1}\left( s\alpha -r\alpha ^{\ast
}\right) -\mathtt{i}\eta _{2}\frac{x}{\mathrm{a}}\right] .  \label{41}
\end{eqnarray}%
Substituting (\ref{40}) and (\ref{41}) into (\ref{38}) we have%
\begin{eqnarray}
&&\left\langle \eta \right\vert U\left( s,r,\kappa =0;\mathrm{a},b=0\right)
\left\vert \eta ^{\prime }\right\rangle  \notag \\
&=&\sqrt{\frac{s}{\mathrm{a}}}\int_{-\infty }^{\infty }\frac{\mathtt{d}x}{2%
\sqrt{\pi }}\exp \left[ -\frac{\eta _{1}^{2}+\eta _{1}^{\prime 2}}{2}+%
\mathtt{i}x(\eta _{2}^{\prime }-\frac{\eta _{2}}{\mathrm{a}})\right]  \notag
\\
&&\times \int \frac{\mathtt{d}^{2}\alpha }{2\pi }\exp \left[ -\frac{s\left(
s^{\ast }-r\right) }{2}\left\vert \alpha \right\vert ^{2}+\left( \eta
_{1}^{\prime }-\eta _{1}r\right) \alpha ^{\ast }\right.  \notag \\
&&\left. +\eta _{1}s\alpha +\frac{1}{4}s\left( r^{\ast }-\allowbreak
s\right) \alpha ^{2}+\frac{1}{4}\left( rs^{\ast }-r^{2}-1\right) \alpha
^{\ast 2}\right]  \notag \\
&=&\sqrt{\frac{\pi }{\mathrm{a}}}\frac{\delta \left( \eta _{2}^{\prime
}-\eta _{2}/\mathrm{a}\right) }{\sqrt{s^{\ast }+r^{\ast }-s-r}}\exp \left[ -%
\frac{1}{2}\left( \eta _{1}^{2}+\eta _{1}^{\prime 2}\right) \right.  \notag
\\
&&\left. +\frac{\allowbreak \left( r^{\ast }-s\right) \eta _{1}^{\prime
2}-\left( s+\allowbreak r\right) \eta _{1}^{2}+\allowbreak 2\eta _{1}\eta
_{1}^{\prime }}{s^{\ast }+r^{\ast }-s-r}\right] .  \label{42}
\end{eqnarray}%
By setting%
\begin{eqnarray}
s &=&\frac{1}{2}\left[ \left( A+D\right) -\mathtt{i}\left( B-C\right) \right]
,  \notag \\
r &=&-\frac{1}{2}\left[ \left( A-D\right) +\mathtt{i}\left( B+C\right) %
\right] ,  \notag \\
s+r &=&D-\mathtt{i}B,\text{ }s-r=A-\mathtt{i}C,\text{\ }s-r^{\ast }=A-%
\mathtt{i}B,  \label{43}
\end{eqnarray}%
which leads to the unimodularity condition for $A,B,C$ and $D,$ i.e., $%
AD-BC=1,$ then we have%
\begin{eqnarray}
&&\left\langle \eta \right\vert U\left( s,r,\kappa =0;\mathrm{a},b=0\right)
\left\vert \eta ^{\prime }\right\rangle  \notag \\
&=&\frac{\pi }{\sqrt{a}}\delta \left( \eta _{2}^{\prime }-\eta _{2}/\mathrm{a%
}\right) \frac{1}{\sqrt{2\mathtt{i}\pi B}}  \notag \\
&&\times \exp \left[ \frac{i}{2B}\left( A\eta _{1}^{\prime 2}-2\eta _{1}\eta
_{1}^{\prime }+D\eta _{1}^{2}\right) \right] .  \label{44}
\end{eqnarray}%
(\ref{44}) shows that $\left\langle \eta \right\vert U\left( s,r,\kappa =0;%
\mathrm{a},b=0\right) \left\vert \eta ^{\prime }\right\rangle $ is just the
kernel of a mixed lens$-$Fresnel transform, in which the variable $\eta _{1}$
of the object experiences an transform of generalized Fresnel transform,
while the variable $\eta _{2}$ undergoes a lens transformation, which
results in a scaling transform.

In sum, based on the entangled-coherent state representation $\left\vert
\alpha ,x\right\rangle $ we have introduced the new kind of continuous WT
which involves both the real variable dilation-transform\ and complex
variable symplectic transform, the new SDWT corresponds to the lens-Fresnel
mixed transform in classical optics. For other WTs, such as a complex WT and
an entangled SWT, we refer to \cite{r18,r19}.

This work is supported by the National Natural Science Foundation of China
(Grant Nos. 10775097 and 10874174). L.-Y. Hu's email address is
hlyun2008@126.com or hlyun@sjtu.edu.cn.

\end{document}